\documentclass[twocolumn,preprintnumbers,amsmath,amssymb]{revtex4}
\usepackage{amstext}
\usepackage{amsmath}
\usepackage{amssymb}
\usepackage{graphicx}
\usepackage{subfigure}

\usepackage{color} 

\begin{document}
\title{Crystallization of hard aspherical particles}

\author{William L. Miller} 
\author{Behnaz Bozorgui}
\author{Angelo Cacciuto}
\email{ac2822@columbia.edu}
\affiliation{Department of Chemistry, Columbia University, 3000 Broadway,\\New York, New York 10027}

\begin{abstract}
We use numerical simulations to study the crystallization of monodisperse systems of  hard aspherical particles.
We find that particle shape and crystallizability can be easily related to each other when particles are
characterized in terms of two simple and experimentally accessible order parameters: one based on the 
particle surface-to-volume ratio, and the other on the angular distribution of the 
perturbations away from the ideal spherical shape. We present a phase diagram obtained by exploring the
crystallizability of  487 different particle  shapes across the two-order-parameter spectrum. Finally, we
consider the physical properties of the crystalline structures accessible to aspherical particles, and 
discuss limits and relevance of our results.
\end{abstract}
\maketitle

\section{Introduction}
Problems of packing and space tiling have fascinated scientists for a very long time.
Kepler's 1611 essay {\it On the Six-cornered Snowflakes} is probably one of the earliest publications on the subject.
Here he conjectured that cubic close packing and hexagonal close packing are the most efficient ways to fill a space using equally sized spheres.
It wasn't until 1998 that Kepler's conjecture was finally announce to be proven (with 99\% degree of confidence) by Thomas Hales~\cite{Hales}.

Recent advances in particle synthesis~\cite{DeVries,Schnablegger,Hong,Weller,Hobbie} have opened the way 
to the production of colloidal particles that are anisotropic both in shape and surface chemistry.
This provides an unlimited number of building blocks that can spontaneously organize into an unprecedented variety 
of structures with potentially novel functional, mechanical, and optical properties.
For these reasons, the problem of efficient packing of nanoparticles is under intense investigation. 

Most of the work on particle crystallization and self-assembly in the last decade  
has  focused on monodisperse~\cite{frenkel1} or polydisperse~\cite{frenkel2,zaccarelli} systems of 
spherical or regularly shaped particles 
(see also~\cite{glotzer,chandler,geissler,cacciuto,torquato,frenkel,glotzer2,glotzer3,glotzer4,esco,weitz} and references therein).
Nevertheless, there are several important cases in which the shape of the single components cannot be tailored at will;
however, an efficient packing, or an understanding of the physical properties of these densely compressed systems, is highly desirable.
Two examples of outstanding problems in this category are the storage of grains~\cite{deGennes} and protein crystallization~\cite{rosenberger}.
Both examples can be ideally thought of as two different aspects of the problem of understanding the role of shape in particle packing.
In the first case the goal is to efficiently pack a system of randomly-shaped polydisperse grains. 
In the second case the aim is to crystallize non-spherical, yet equally shaped monodisperse components. 
The latter case is the focus of the present paper.  

We want to understand  how the ability of 
particles to form macroscopically ordered crystal structures is affected by their shape. Specifically, we
analyze how random perturbations from the ideal spherical shape affect the crystallizability of 
a densely packed system of indistinguishable hard particles. Although a few papers have  dealt with the thermodynamic behavior of 
soft/deformable particles as a model for polymer brushes or polymer-coated colloids 
(\cite{pamies,capone,richter,bozorgui,ziherl} and references therein), 
to the best of our knowledge, this paper is the first  study 
where shape distortions, frozen onto the particles, are explicitly and systematically accounted for.

The problem of crystal formation is poorly understood. From a thermodynamic standpoint, 
classical nucleation theory informs us that a crystal can only be formed via a barrier crossing event.
The free energy gain to form a nucleus of a stable crystalline structure in a supersaturated solution 
has to balance out the free energy cost associated with the formation of an interface between the solid and the fluid parent phase.
The Gibbs free energy cost, $\Delta G$, associated with this event has a strong dependence on the interfacial free energy, $\gamma$:
${\Delta}G \propto \frac{\gamma^3}{(\rho\Delta\mu)^2}$,  where $\rho$ and $\Delta\mu$ 
are the density and the chemical potential difference between solid and fluid phase, respectively.
As $\gamma$ is very hard to extract from experiments, computer simulations are
the tool of choice to study and analyze this important process in detail.

One way to investigate our problem could be 
to measure the dependence of  $\gamma$ (or $\Delta G$) on the magnitude of the shape perturbations at constant $\Delta\mu$.
This approach would be appropriate when considering a system that is polydisperse in shape and size; however, in our case,
we have to study a statistically relevant number of systems  each containing $N$ identical particles with the 
same shape perturbation. Such an approach is therefore impractical because (a) it would require the 
computation of a phase diagram for every system with a given particle realization (about 500 in this study)
and (b) for sufficiently large shape deformations the FCC structure that nucleates out of a solution of hard spheres may 
not  necessarily be the most stable one, and this would lead to a meaningless comparison of unrelated $\gamma$s.

Given these limitations, our strategy will be to generate a large number of particle shapes, sort them in terms of the extent 
of their asphericity, and simply analyze whether they organize into an ordered 
and periodic structure (whatever that might be) once compressed at large densities.
We are left with the problem of devising a new
model that allows for an effective control of the particle's degree of asphericity. 
Unfortunately there is not a unique way of introducing perturbations on the particle shape, so we have opted for a model that,
without loss of generality, weds simplicity and numerical efficiency. 

\section{Model}
In our model, each particle is built by setting the center of $N_b$ ($4 \leq N_b \leq 12$) spheres of diameter $\sigma$ {at random positions}
inside a spherical shell of diameter  $\sigma_0<\sigma$. The overall volume generated from the resulting 
overlapping aggregate defines our new particle. Deviations from the ideal spherical shape 
can be conveniently controlled by varying  $\sigma_0$ and $N_b$.  {Values of $\sigma_0$ and $N_b$ were chosen to achieve wide coverage of the range of possible particle geometries.}
For  $\sigma_0=0$ one recovers the spherical limit, and as  $\sigma_0$ increases, 
particles develop larger and larger shape distortions. In a similar fashion, large values of $N_b$ 
result in a bumpy but overall isotropic particle, whereas small values of $N_b$ tend to generate very anisotropic shapes.
Once a particle is built, the center of mass of this cluster of balls is determined, 
and the entire cluster is scaled so that its total volume equals that of a spherical particle of diameter $\sigma$, i.e. $\frac{\pi}{6}\sigma^3$.
Any two particles $i$ and $j$ interact via a hard repulsive potential defined as
\begin{equation}
U_{ij}=
\begin{cases}
0 & {\textrm{if }} |r_s-r_t|>\sigma_R \,\,\,\,\,\,\forall s\in i \,\,,\,\, \forall t\in j \cr
\infty &\text{otherwise}\cr
\end{cases}
\end{equation}
where $s$ and $t$ run over all spheres of rescaled diameter $\sigma_R$ constituting particle $i$ and particle $j$ respectively.
Experimental realizations of colloidal particles similar to ours could be generated using the approach described in reference~\cite{weitz} to create uniform nonspherical particles with tunable shapes.  

Figure 1 shows a few snapshots of particle shapes obtained for different values of $N_b$ and  $\sigma_0$. 
\begin{figure}
	\includegraphics[width=0.36\textwidth]{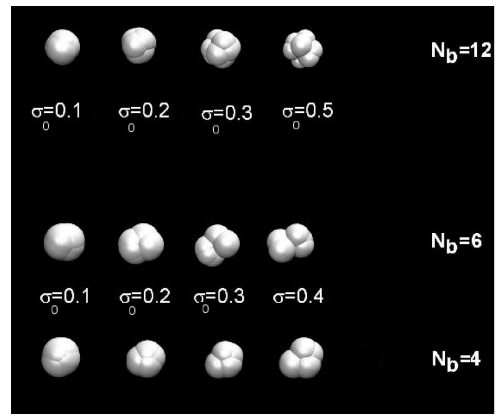}
	\caption{Model particles for different values of $\sigma_0$ and $N_b$ built according to the scheme described in the text.}
\end{figure}

To analyze the high density behavior of these systems we used Monte Carlo simulations in the $NPT$ ensemble.
Given the large number of simulations involved in this study, we limited the size of our systems to a total 
of $N=128$ particles. {We use a cubic box with periodic boundary conditions.}  Using as a reference the hard-spheres model, i.e. coexistence reduced pressure and volume fraction respectively at $P^*\simeq 11.6$ and 
$\phi\simeq 0.494$, we performed, for each system with a given particle realization, a series of simulations at increasing values of pressure beginning at $P^*=10$ with a 
typical increment of $\Delta P^*$=0.5 or smaller. We ran each simulation for a minimum of $4\cdot 10^6$ Monte Carlo sweeps after thermalization. 
The largest pressure was set by either the crystallization of the system, or by the reaching of a volume fraction $\phi\simeq 0.6$, above the 
glass transition point of hard spheres $\phi_G\simeq 0.58$. Crystallization was detected in our samples with a combination of methods:
(a) standard spherical-harmonics based bond order parameter $q_6$~\cite{q61,q62}, (b) a careful monitoring 
of the system volume fraction over time for sudden jumps, and (c) visual inspection. We investigated a total of 487 different particle shapes.

We find that at least two order parameters are required to properly characterize the shape of each particle. 
The first is its asphericity  $A$ (as opposed to the commonly used sphericity, $S$~\cite{sphericity}), defined in terms of the 
surface to volume ratio of a particle $\alpha_p=A_p/V_p$ with respect to that of a sphere of diameter $\sigma$, $\alpha_s=6/\sigma$, as
\[A = 1 - S = 1-\frac{\alpha_s}{\alpha_p}.\]
Given our model setup, $V_p=V_s$, $A$ simplifies to $A=1-\pi\sigma^2/A_p$ 
 
The second parameter, $q$, measures the orientational symmetry  of the particle.
It is used to describe the  asphericity of random walks~\cite{rudnick}, and it is 
obtained by combining invariants of the particle inertia tensor $I_{ij}$  as
\[q=\frac{\left(R_1^2-R_2^2\right)^2 + \left(R_1^2-R_3^2\right)^2 + \left(R_2^2-R_3^2\right)^2}{2\left(R_1^2 + R_2^2 + R_3^2\right)},\]
where $R_1$, $R_2$, and $R_3$ are the three principal eigenvalues of the inertia tensor of the particle, i.e., 
the three principle radii of gyration of the particle.
Both parameters are defined so that they are equal to $0$ for a perfectly spherical particle,
and approach $1$ for  extremely aspherical ones. 
Note that $A$ depends intimately on the value of $\sigma_0$ used to construct the particle, whereas $q$ depends only on 
the angular distribution of spheres about the center of mass -- that is, it is completely independent of $\sigma_0$.

\section{Results}
\begin{figure}
	\includegraphics[width=0.5\textwidth]{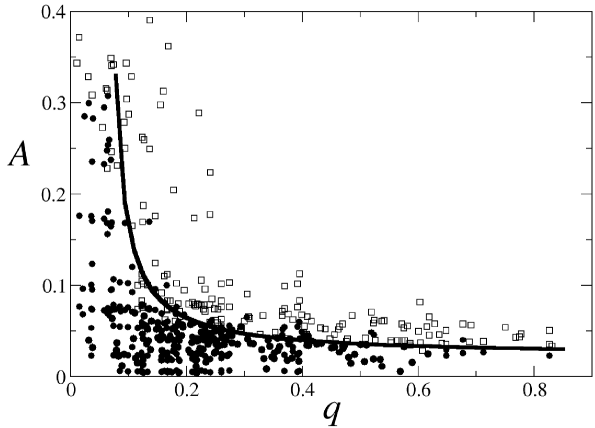}
	\caption{Crystallizability of aspherical particles characterized in terms of two shape parameters 
    $A$ and $q$. Filled circles indicate particles that {easily} crystallized, while open squares indicate particles that did not.  
{Data include results for 487 total particle geometries, including 85 using $N_b=4$, 200 using $N_b=6$, 92 using $N_b=8$, and 110 using $N_b=12$.}  The solid line is a guide to the eye.}\label{plot}
\end{figure}

Figure~\ref{plot} summarizes the main result of our paper. 
It is obtained by collecting 
the crystallizability of the 487 systems built out of the 487
different particle geometries we have generated across the $A$, $q$ spectrum. 
Each point in the $A$ vs $q$ diagram represents the result of a set of simulations at different pressures.
As would be expected, crystallization is favored when both $A$ and $q$ are small $-$ that is, 
when the particles are nearly spherical by both measures.  
A roughly inverse relationship is clearly evident; particles with large $A$ must have very small $q$ in order to have
a hope of crystallization, and vice-versa.
But more importantly, we find the existence of a clear boundary delineating the crystallizability limit 
for every possible shape generated with our model (small deviations at the interface are
likely due to finite size effects and or the limited length of our simulations).
This is quite remarkable because it provides a very useful way of predicting whether a particular particle shape can pack into a
crystalline structure by simply measuring the experimentally accessible $A$ and $q$. 

We find that a good empirical expression to describe the phase boundary is  $A(q) = 0.023 + 1/(170q-10)$. This curve
is only  meant to serve as a guide to the eye and to give an approximate numerical estimate of the 
location of the phase boundary for $q\gtrsim 0.08$. Notice that for smaller values of $q$ our data seem to 
indicate a sharp end of the crystal boundary. We believe this to be an artifact of our particle model.
In fact, that region is where the value of $\sigma_0$ becomes sufficiently  large ($\sigma_0>0.5\sigma$)
to break the compactness of the particles generated with our method, especially those with smaller values of $N_b$. 
Furthermore, as small values of $q$ indicate a large orientational symmetry, we expect this
region to be heavily populated by specific  geometric arrangements, whose packing properties, at large values of $A$,
will be extremely sensitive of the particular value of $N_b$. 

As our model is intended to describe randomly shaped particles, 
our diagram does not include the results for particles designed with very specific shapes such as rods,
plates or regular polyhedric geometries that are known to crystallize. 
These particular cases would generate sharp peaks around specific values of $q$. 
Furthermore, is not clear that our two order parameters, that have after all been  selected to describe asphericity, 
would be the most appropriate to study deviations from an arbitrary non spherical designed shape. 
We therefore limited our study to $N_b >3$, to explicitly avoid trivial cases as rod-like ($N_b=2$) and plate-like particles ($N_b=3$).

Two obvious questions present themselves in the face of these data.
The first is: what is the nature of the crystals formed when particles do {easily} crystallize; 
does the system present translational but not orientational order, as expected for $A\simeq 0$, or
does the rotational motion of the particles becomes restricted for large values of $A$?
The second is: what sets the boundary between the two phases;
do the particles that fail to {easily} crystallize do so because they become kinetically trapped or because of the 
lack of a stable crystal phase?
\begin{figure}
	\subfigure{
		\includegraphics[width=0.4\textwidth]{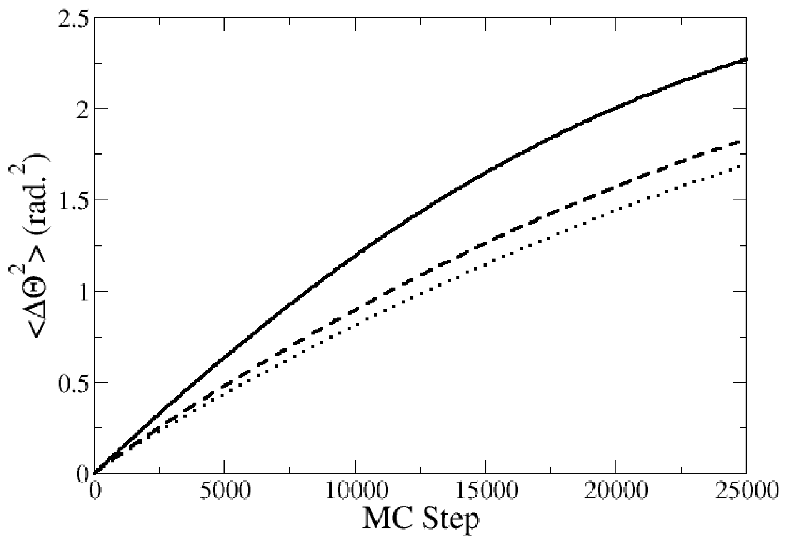}
		\put(-200,140){\bf{(a)}}
		\label{rot}
	}
	\subfigure{
		\includegraphics[width=0.4\textwidth]{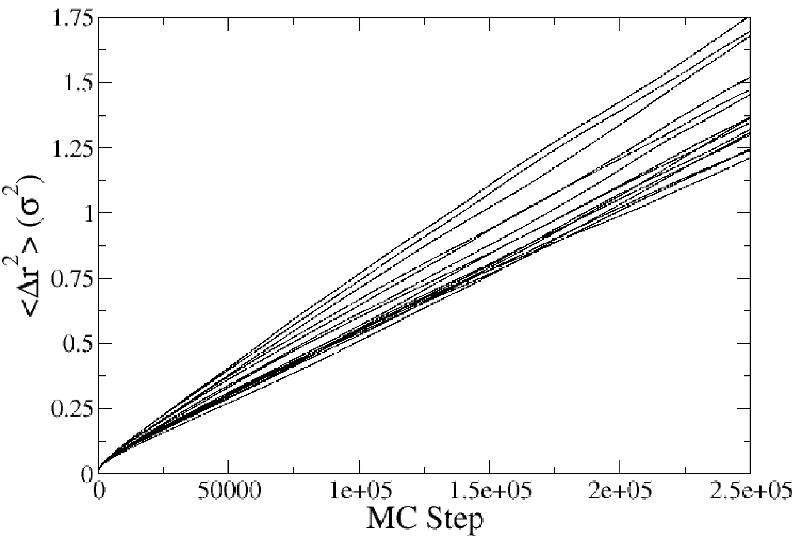}
		\put(-200,140){\bf{(b)}}
		\label{trans}
	}
	\caption{(a) Rotational mean square displacement for a subset of the particles considered. The solid line is a 
	reference for nearly spherical particles. The dashed line is the result for  those particles that  
	 {easily} crystallized and the dotted line shows $\left<\Delta\theta^2\right>$ for those that did not.
  The data were averaged over 20 realizations in each region under the same pressure $P^*=20$.
(b) Translational mean square displacement for  20 different particle shapes that failed to {easily} crystallize.}\label{diff}
\end{figure}

To address the first question, we measured the rotational diffusion, $\left<\Delta\theta^2\right>$, for 20 systems 
that {easily} crystallize, all at a reduced pressure $P^* = 20$ and near the phase boundary.  As a reference 
we also plotted  $\left<\Delta\theta^2\right>$ for nearly spherical particles, obtained by setting $\sigma_0 = 10^{-4}\sigma$, 
where we know particles are free to rotate at their lattice sites.
As can be seen in Fig.~\ref{rot}, we find no evidence that particles in the crystalline phase become 
orientationally arrested or manifest an orientationally anomalous behavior. The only effect is that of decreasing their diffusion constant, 
but this is expected from simple geometrical considerations. It is obvious that at very large densities, 
regardless of the specific phase a system selects, particles' orientations will manifest a glassy behavior or eventually freeze.~\cite{schweitzer} 
It is therefore of interest to also look at the dynamical
properties of those particles in systems do not {easily} crystallize and are located 
just across the phase boundary from the ones that do {easily} crystallize. 
These results, obtained at the same reduced pressure $P^*=20$, are also shown in Fig.~\ref{rot}. 
We find no signature of anomalous dynamics, neither in the rotational (Fig.~\ref{rot}), nor in the translational 
(Fig.~\ref{trans}) degrees of freedom, i.e. such systems behave as regular fluids.

This result is by no means conclusive, as a thorough investigation of this last point would require larger system sizes 
and an event-driven dynamics of the components.  Our results seem to suggest that what sets the location of 
the phase boundary is not a sudden slowdown of the dynamics of the system, but more likely
 an increase of the Gibbs free energy difference between the crystalline  and the fluid phase, analogously to
that found for polydisperse spherical particles~\cite{frenkel2}. 
It would be interesting to investigate the equilibrium properties and the stability of 
candidate crystalline structures for different values of $A$ and $q$ to investigate whether our boundary line 
coincides with the onset of crystal instability, but for the reasons given above, we have not attempted to do so.

\section{Conclusion}
Apart from the details concerning the dynamics of these systems, our results show that, for aspherical particles,
shape and crystallizability can be directly and easily correlated when particles are characterized in terms of their
asphericity via $q$ and $A$. 
{It is possible that for some highly asymmetric particle geometries a non-cubic box would accommodate crystallization whereas a cubic one would not; however, we believe this is a secondary effect and its relevance is likely within the range of errors due to finite size effects.}
Our data suggest precise limits for the manufacture of nanocomponents expected to crystallize,
and may have important implications for the problem {of} protein crystallization. 
The latter system is clearly far more complex  than the one we explored,
since not only shape, but also interparticle direct interactions (which are not necessarily isotropic), 
are responsable for the organization of proteins into large macroscopic crystals. Nevertheless, it would be
interesting to systematically explore to what extent  a similar correlation exists in this case.   

\section*{Acknowledgments}
This work was supported by the National Science Foundation
under CAREER Grant No. DMR-0846426.

\end{document}